\documentclass[a4paper,11pt]{article}

\usepackage{times}

\usepackage[english]{babel}
\usepackage{amsmath,amssymb,url}
\usepackage[dvips]{graphics}
\normalsize
\begin{document}

\def\csbul{\leavevmode{\rule[0.14em]{0.24em}{0.24em}}\kern 0.33em}
\def\TitleSkip{0mm}
\def\AbstractLeftMargin{0mm}

\makeatletter
\renewenvironment{thebibliography}[1]
     {\section{\refname}%
      \@mkboth{\refname}{\refname}
      \normalfont
      \list{\@biblabel{\@arabic\c@enumiv}}%
           {\settowidth\labelwidth{\@biblabel{#1}}%
            \leftmargin\labelwidth
            \advance\leftmargin\labelsep
            \parsep -2.5\p@ \@plus\p@ \@minus\p@
            \@openbib@code
            \usecounter{enumiv}%
            \let\p@enumiv\@empty
            \renewcommand\theenumiv{\@arabic\c@enumiv}}%
      \sloppy
      \clubpenalty4000
      \@clubpenalty \clubpenalty
      \widowpenalty4000%
      \sfcode`\.\@m}
     {\def\@noitemerr
       {\@latex@warning{Empty `thebibliography' environment}}%
      \endlist}
\makeatother




\title{General relativity and cosmology derived from principle of maximum power or force}

\author{Christoph Schiller\footnote{Giesecke \& Devrient Research\&Development, Prinzregentenstra{\ss}e, 81677
M\"unchen, Germany, \rm {christoph.schiller@motionmountain.org}.}}

\date{\today}

\maketitle

\normalfont\normalsize

\abstract{\noindent The field equations of general relativity are shown to
derive from the existence of a limit force or of a limit power in nature.  The
limits have the value of $c^4/4G$ and $c^5/4G$.  The proof makes use of a
result by Jacobson.  All known experimental data is consistent with the
limits.  Applied to the universe, the limits predict its darkness at night and
the observed scale factor.  Some experimental tests of the limits are
proposed.  The main counter-arguments and paradoxes are discussed, such as the
transformation under boosts, the force felt at a black hole horizon, the
mountain problem, and the contrast to scalar--tensor theories of gravitation.
The resolution of the paradoxes also clarifies why the maximum force and the
maximum power have remained hidden for so long.  The derivation of the field
equations shows that the maximum force or power plays the same role for
general relativity as the maximum speed plays for special relativity.

\medskip


\medskip

}



\newpage\normalfont\normalsize
\section{Introduction}

\noindent A simplification of general relativity has welcome effects on the
teaching of the topic.  Some years ago, a strong simplification has been
reported by Gibbons (Gibbons 2002) and independently, by the present author
(Schiller 1997-2004) General relativity was shown to derive from the
so-called maximum force (or maximum power) principle: \emph{There is a maximum
force (and power) in nature:}
\begin{equation}
    F \leq \frac{c^4}{4G} = {3.0\cdot 10^{43}}{\rm N}\quad
    \quad\hbox{and}\quad\quad P \leq \frac{c^5}{4G} = {9.1\cdot 10^{51}}{\rm
    W}\quad.
    \label{ffff}
\end{equation}
Either of the two equivalent expressions can be taken as basic principle.  So
far, the arguments used for the connection between these limits and general
relativity were either quite abstract 
or rather
heuristic. 
The present paper gives a derivation of the field equations from either of the
limit values and shows the equivalence of the two formulations of general
relativity.  This paper also includes the discussion of the main paradoxes,
uses the limits (\ref{ffff}) to deduce the central points of cosmology and
suggests some new experimental tests of general relativity.

The concept of force needs careful use in general relativity.  Force is the
change of momentum with time.  Since momentum is a conserved quantity, force
is best visualized as an upper limit for the rate of flow of momentum (through
a given physical surface).  Only with this clarification does it make sense to
use the concept of force in general relativity.

The value $c^4/4G$ of the force limit is the energy of a Schwarzschild black
hole divided by twice its radius.  The maximum power $c^5/4G$ is realized when
such a black hole is radiated away in the time that light takes to travel
along a length corresponding to twice the radius.  It will become clear below
why a Schwarzschild black hole, as an extremal case of general relativity, is
necessary to realize these limit values.

Generally speaking, the aim is to prove that the principle of maximum force
(or that of maximum power) plays for general relativity the same role that the
principle of maximum speed plays for special relativity.  This unconventional
analogy (Schiller 1997-2004) 
requires a proof in several steps.  First, one has to
\emph{derive} the field equations of general relativity from the maximum
force.  Then one has to show that \emph{no imaginary} set-up or situation --
thus no Gedanken experiment -- can overcome the limit.  Subsequently, one has
to show that \emph{no experimental data} contradicts the statement of maximum
force.  Finally, one has to deduce predictions for future experimental tests
made on the basis of maximum force or power.  These are the same steps that
have lead to the establishment of the idea of a maximum speed of nature.
These steps structure this paper.

Since force (respectively power) is change of momentum (energy) with time, the
precise conditions for momentum (energy) measurement must be specified, in a
way applicable in a space-time that is curved.  Momentum, like energy, is a
conserved quantity.  Any change of momentum or energy thus happens through
flow.  As a result, a maximum force (respectively power) value in nature
implies the following statement: one imagines a physical surface and
completely covers it with observers; then the integral of all momentum
(respectively energy) values flowing through that surface per time, measured
by all those observers, never exceeds the maximum value.  It plays no role how
the surface is chosen, as long as it is \emph{physical}, i.e., as long as the
surface allows to fix observers on it.

A condition for such a measurement is implicit in the surface flow
visualization.  The local momentum (or energy) change for each observer is the
value that each observer measures for the flow at precisely his or her
position.  The same condition of observer proximity is also required for speed
measurements in special relativity.

Since 3-force and power appear together in the force 4-vector, both the force
and the power limits are equivalent and inseparable.  It is sometimes
suggested that the theory of general relativity does not admit a concept of
force or of its zeroth component, power.  This is not correct; the value of
force is simply so strongly dependent on observer choices that usually one
tends to avoid the concept of force altogether.  On the contrary, it turns out
that \emph{every} quantity with the dimensions of force (or of power) that is
measured by an observer is bound by the limit value.  This can either be the
magnitude of the four vector or the value of any of its four components.  The
force limit $c^4/4G$ (and the corresponding power limit) is valid for all
these observables, as will become clear below.  In particular, it will be
shown below why an arbitrary change of coordinates does \emph{not} allow to
exceed the force or power limit, contrary to expectation.  This result of
general relativity is equivalent to the result of special relativity that a
change of coordinates does not allow to exceed the speed limit.

The maximum force and power are also given, within a factor 1/4, by the Planck
energy divided by the Planck length, respectively, by the Planck time.  The
origin of the numerical coefficient $1/4$ has no deeper meaning.  It simply
turns out that $1/4$ is the value that leads to the correct form of the field
equations of general relativity.

\section{The derivation of general relativity}

\noindent To derive the theory of relativity one has to study those systems
that realize the limit value of the observable under scrutiny.  In the the
case of the \emph{special} theory of relativity, the systems that realize the
limit speed are light and massless particles.  In the \emph{general} theory of
relativity, the systems that realize the limit are less obvious.  One notes
directly that a maximum force (or power) cannot be realized across a
\emph{volume} of space.  If that were the case, a simple boost could transform
the force (or power) to a higher value.  Nature avoids this by realizing
maximum force and power only on surfaces, not volumes, and at the same time by
making such surfaces unattainable.  These unattainable surfaces are basic to
general relativity; they are called \emph{horizons}.  Maximum force and power
only appears on horizons (Schiller 1997-2004). 
The definition of a horizon as a surface
of maximum force (or power) is equivalent to the more usual definition as a
surface that provides a limit to signal reception, i.e., a limit to
observation.

The reasoning in the following will consists of three additional steps.
First, it will be shown that a maximum force or power implies that
unattainable surfaces are always curved.  Then it will be shown that any
curved horizon follows the so-called horizon equation.  Finally, it will be
shown that the horizon equation implies general relativity.  (In fact, the
sequence of arguments can also be taken in the opposite direction; all these
steps are equivalent to each other.)

The connection between horizons and the maximum force is the central
point in the following.  It is as important as the connection between
light and the maximum speed in special relativity.  In special
relativity, one shows that light speed, being the maximum speed in
nature, implies the Lorentz transformations.  In general relativity,
one must show that horizon force, being the maximum force in nature,
implies the field equations.  To achieve this aim, one starts with the
realization that all horizons show energy flow at their location.
There is no horizon without energy flow. 
This connection implies that a horizon cannot be a plane, as an infinitely
extended plane would imply an infinite energy flow.

The simplest finite horizon is a static sphere.  A spherical horizon is
characterized by its curvature radius $R$ or equivalently, by its surface
gravity $a$; the two quantities are related by $2aR=c^2$.  The energy flow
moving though any horizon is always finite in length, when measured along the
propagation direction.  One can thus speak more specifically of an energy
pulse.  Any energy pulse through a horizon is thus characterized by an energy
$E$ and a proper length $L$.  When the energy pulse flows perpendicularly
through a horizon, the momentum change or force for an observer at the horizon
is
\begin{equation}
    F = \frac{E}{L} \quad.
    \end{equation}
The goal is to show that maximum force implies general relativity.  Now, the
maximum force is realized on horizons.  One thus needs to insert the maximum
possible values for each of these quantities and to show that general
relativity follows.

Using the maximum force value and the area $4 \pi R^2$ for a spherical horizon
one gets
\begin{equation}
    \frac{c^4}{4G} = \frac{E }{LA} 4 \pi R^2 \quad.
    \label{s5656}
    \end{equation}
The fraction $E/A$ is the energy per area flowing through any area $A$ that is
part of a horizon.  The insertion of the maximum values is complete when one
notes that the length $L$ of the energy pulse is limited by the radius $R$.
The limit $L\leq R$ is due to geometrical reasons; seen from the concave side
of the horizon, the pulse must be shorter than the curvature radius. 
An independent argument is the following. 
The length $L$ of an object accelerated by $a$ is limited by special
relativity (D'Inverno 1992, Rindler 2001)  
by 
\begin{equation}
    L \leq \frac{c^2 }{2a} \quad.
\end{equation}
Special relativity already shows that this limit is due and related to the
appearance of a horizon.  Together with relation (\ref{s5656}), the statement
that horizons are surfaces of maximum force leads to the following central
relation for static, spherical horizons:
\begin{equation}
    E = \frac{c^2}{8 \pi G} \, a\, A \quad.
    \label{b191203}
\end{equation}
This \emph{horizon equation} relates the energy flow $E$ through an area $A$
of a spherical horizon with surface gravity $a$.  The horizon equation follows
from the idea that horizons are surfaces of maximum force.  The equation
states that the energy flowing through a horizon is limited, that this energy
is proportional to the area of the horizon, and that the energy flow is
proportional to the surface gravity.

The above derivation also yields the intermediate result
\begin{equation}
    E \leq \frac{c^4}{16 \pi G} \, \frac{A}{L} \quad.
    \label{huhjilko}
\end{equation}
This form of the horizon equation states more clearly that no surface other
than a horizon can reach the limit energy flow, given the same area and pulse
length (or surface gravity).  No other part of physics makes comparable
statements; they are an intrinsic part of the theory of gravitation.

Another variation of the derivation of the horizon starts with the
emphasis on power instead of on force.  Using $P=E/T$ as starting
equation, changing the derivation accordingly, also leads to the
horizon equation.  

It is essential to stress that the horizon equations (\ref{b191203}) or
(\ref{huhjilko}) follow from only two assumptions: first, there is a maximum
speed in nature, and second, there is a maximum force (or power) in nature.
No specific theory of gravitation is assumed.  The horizon equation might even
be testable experimentally, as argued below.  (One also notes that the horizon
equation -- or, equivalently, the force or power limits -- imply a maximum
mass change rate in nature given by $dm/dt \leq c^3/4G$.)  In particular, up
to this point it was \emph{not} assumed that general relativity is valid;
equally, it was \emph{not} assumed that spherical horizons yield Schwarzschild
black holes (indeed, other theories of gravity also lead to spherical
horizons).

Next one has to generalize the horizon equation from static and spherical
horizons to general horizons.  Since the maximum force is assumed to be valid
for \emph{all} observers, whether inertial or accelerating, the generalization
is straightforward.  For a horizon that is irregularly curved or time-varying
the horizon equation becomes
\begin{equation}
    \delta E = \frac{c^2}{8\pi G} \, a\, \delta A \quad.
    \label{bb191203}
\end{equation}
This differential relation -- it might be called the \emph{general horizon
equation} -- is valid for any horizon.  It can be applied separately for every
piece $\delta A$ of a dynamic or spatially changing horizon.  The general
horizon equation~(\ref{bb191203}) is known to be equivalent to general
relativity at least since 1995, when this equivalence was implicitly given
by Jacobson (Jacobson 1995). 
It will be shown that the differential horizon
equation has the same role for general relativity as $dx=c\;dt$ has for
special relativity.  From now on, when speaking of the horizon equation, the
general, differential form (\ref{bb191203}) of the relation is implied.
  
It is instructive to restate the behaviour of energy pulses of length $L$ 
in a way that holds for any surface, even one that is not a horizon. 
Repeating the above derivation, one gets
\begin{equation}
    \frac{\delta E}{\delta A} \leq \frac{c^4}{16\pi G} \, \frac{1}{L} \quad.
     \label{b64745747}
\end{equation}
Equality is only reached in the case that the surface $A$ is a horizon.  In
other words, whenever the value ${\delta E}/{\delta A}$ approaches the right
hand side, a horizon is formed.  This connection will be essential in the
discussion of apparent counter-examples to the limit values.

If one keeps in mind that on a horizon, the pulse length $L$ obeys $L \leq
c^2/2a$, it becomes clear that the general horizon equation is a consequence
of the maximum force $c^4/4G$ or the maximum power $c^5/4G$.  In addition, the
horizon equation takes also into account maximum speed, which is at the origin
of the relation $L \leq c^2/2a$.  The horizon equation thus follows purely
from these two limits of nature.  One notes that one can also take the
opposite direction of arguments: it is possible to derive the maximum force
from the horizon equation (\ref{bb191203}).  The two statements are thus
equivalent.

One notes that the differential horizon equation is also known under the name
`first law of black hole mechanics' (Wald 1993).  
The arguments so far thus show that the first law of black hole mechanics is a
consequence of the maximum force or power in nature.  This connection does not
seem to appear in the literature so far.  The more general term `horizon
equation' used here instead of `first law' makes three points: first, the
relation is valid for any horizon whatsoever; second, horizons are more
fundamental and general entities than black holes are; third, horizons are
limit situations for physical surfaces.

The remaining part of the argument requires the derivation of the field
equations of general relativity from the general horizon equation.  The
derivation -- in fact, the equivalence -- was implicitly provided by Jacobson
(Jacobson 1995), 
and the essential steps are given in the following.  (Jacobson did not stress
that his derivation is valid also for continuous space-time and that his
argument can also be used in classical general relativity.)  To see the
connection between the general horizon equation (\ref{bb191203}) and the field
equations, one only needs to generalize the general horizon equation to
general coordinate systems and to general directions of energy-momentum flow.
This is achieved by introducing tensor notation that is adapted to curved
space-time.

To generalize the general horizon equation, one introduces the
general surface element $d \Sigma$ and the local boost Killing vector
field $k$ that generates the horizon (with suitable norm).  Jacobson
uses the two quantities to rewrite the left hand side of the
general horizon equation (\ref{bb191203}) as
\begin{equation}
    \delta E = \int T_{ab} k^a d\Sigma^b \quad,
\end{equation}
where $T_{ab}$ is the energy-momentum tensor.  This expression
obviously gives the energy at the horizon for arbitrary coordinate
systems and arbitrary energy flow directions.
	
Jacobson's main result is that the the right hand side of the
general horizon equation (\ref{bb191203}) can be rewritten,
making use of
the (purely geometric) Raychaudhuri equation, as
\begin{equation}
    a \; \delta A = {c^2} \int R_{ab} k^a d\Sigma^b \quad,
\end{equation}
where $R_{ab}$ is the Ricci tensor describing space-time curvature.  This
relation thus describes how the local properties of the horizon depend on the
local curvature.  One notes that the Raychaudhuri equation is a purely
geometric equation for manifolds, comparable to the expression that links the
curvature radius of a curve to its second and first derivative.  In
particular, the Raychaudhuri equation does \emph{not} contain any implications
for the physics of space-times at all.

Combining these two steps, the general horizon equation
(\ref{bb191203}) becomes
\begin{equation}
    \int T_{ab}k^a d\Sigma^b = \frac{c^4}{8\pi G} \int R_{ab}k^a
    d\Sigma^b \quad.
    \label{c191203}
\end{equation}
Jacobson then shows that this equation, together with local
conservation of energy (i.e., vanishing divergence of the
energy-momentum tensor), can only be satisfied if
\begin{equation}
    T_{ab} = \frac{c^4}{8\pi G} \left (R_{ab}-(\frac{R}{2}+\Lambda)
    g_{ab} \right ) \quad,
    \label{d191203}
\end{equation}
where $R$ is the Ricci scalar and $\Lambda$ is a constant of integration whose
value is not specified by the problem.  These are the full field equations of
general relativity, including the cosmological constant $\Lambda$.  The field
equations thus follow from the horizon equation.  The field equations are
therefore shown to be valid at horizons.

Since it is possible, by choosing a suitable coordinate transformation, to
position a horizon at any desired space-time event, the field equations must
also be valid over the whole of space-time.  This conclusion completes the
result by Jacobson.  Since the field equations follow, via the horizon
equation, from maximum force, one has thus shown that at every event in nature
the same maximum possible force holds; its value is an invariant and a
constant of nature.

The reasoning shown here consisted of four steps.  First, it was shown that a
maximum force or power implies the existence of unattainable surfaces, which
were called horizons.  Second, is was shown that a maximum force or power
implies that unattainable surfaces are always curved.  Third, it was shown
that any curved horizon follows the horizon equation.  Forth, it was shown (in
the way done by Jacobson) that the horizon equation implies general
relativity.

In other words, the field equations of general relativity are a direct
consequence of the limited energy flow at horizons, which in turn is due to
the existence of a maximum force (or power).  In fact, the argument also works
in the opposite direction, since all intermediate steps are equivalences.
This includes Jacobson's connection between the horizon equation and the field
equations of general relativity.  Maximum force (or power), the horizon
equation, and general relativity are thus \emph{equivalent}.  As a result, one
finds the corollary that \emph{general relativity implies a maximum force}.

The maximum force (or power) has thus the same double role in general
relativity that the maximum speed has in special relativity.  In
special relativity, the speed of light is the maximum speed; at the
same time it is the proportionality constant that connects space and
time, as in $dx=c\;dt$.  In general relativity, the horizon force is
the maximum force; at the same time the maximum force appears (adorned
with a factor $2\pi$) in the field equations as the proportionality
constant connecting energy and curvature.  If one prefers, the maximum
force thus describes the elasticity of space-time and at the same time
it describes -- if one dares to use the simple image of space-time as
a medium -- the maximum tension to which space-time can be subjected.
This double role of material constants both as proportionality factor and
as limit value is well-known in material science.

The analogy between special and general relativity can be carried further.  In
special relativity, maximum speed implies $dx=c\;dt$ and the observation that
time changes with observer change.  In general relativity, maximum force (or
power) imply the horizon equation 
$\delta E = a\, \delta A \, {c^2}/{8\pi G}$ 
and the observation that space-time is curved.  Curvature is a result of
the maximum force or maximum power.  Indeed, the derivation above showed that
a finite maximum force implies horizons that are curved; the curvature of
horizons imply the curvature of surrounding space--time.

One might ask whether rotating or charged black holes change the argument
that lead to the derivation of general relativity.  However, the derivation
using the Raychaudhuri equation does not change.  In fact, the only change of
the argument appears with the inclusion of torsion, which changes the
Raychaudhuri equation itself.  As long as torsion plays no role, the
derivation given above remains valid. 

Another question is how the above proof relates to scalar--tensor theories of
gravity.  If a particular scalar-tensor theory would obey the general horizon
equation (\ref{bb191203})
%
%
then it would also show a maximum force.  The general horizon equation must be
obeyed both for \emph{static} and for \emph{dynamic} horizons.  If that is the
case, the specific scalar--tensor theory would be equivalent to general
relativity, as it would allow, using the argument of Jacobson, to deduce the
usual field equations.  This case can appear if the scalar field behaves like
matter, i.e., if it has mass-energy like matter and curves space-time like
matter.  On the other hand, if in the particular scalar--tensor theory the
general horizon equation (\ref{bb191203}) is not obeyed for \emph{all moving}
horizons -- which is the general case, as scalar--tensor theories have more
defining constants than general relativity -- then the maximum force does not
appear and the theory is not equivalent to general relativity.  This
connection also shows that an experimental test of the horizon equation for
\emph{static} horizons only is not sufficient to confirm general relativity;
such a test rules out only some, but not all scalar--tensor theories.

\section{Apparent counter-arguments and paradoxes}

Despite the preceding and others proofs (Gibbons 2002) 
for the equivalence of
maximum force and the equations of general relativity, the idea of a maximum
force is not yet common.  Indeed, maximum force, maximum power and maximum
mass change directly induce counter-arguments and attempts to exceed the
limit.

\begin{figure}[t]%
  \hbox to \textwidth{\hss\includegraphics{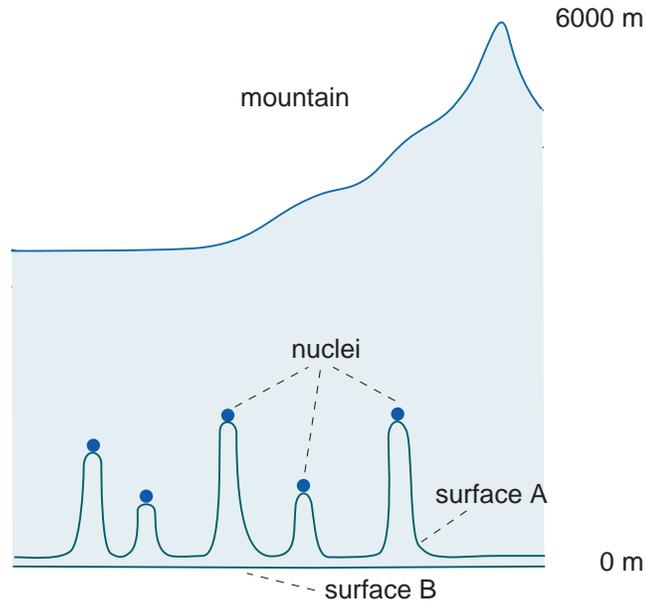}\hss}%
  \medskip
  \caption{The mountain problem}%
  \protect\label{mofig}%
\end{figure}

\csbul The mountain attempt.  It is possible to define a surface that is so
strangely bent that it passes \emph{just below} every nucleus of every atom of
a mountain, like the surface A in Figure~\ref{mofig}.  All atoms of the
mountain above sea level are then \emph{just above} the surface, barely
touching it.  In addition, one imagines that this surface is moving
\emph{upwards} with almost the speed of light.  It is not difficult to show
that the mass flow through this surface is higher than the mass flow limit.
Indeed, the mass flow limit $c^3/4G$ has a value of about $10^{35}$\;kg$/$s;
in a time of $10^{-22}$\;s, the diameter of a nucleus divided by the speed of
light, only $10^{13}$\;kg need to flow through the surface; that is the mass
of a mountain.

The mentioned surface seems to provide a counter-example to the limit.
However, a closer look shows that this is not the case.  The issue is the
expression ``just below".  Nuclei are quantum particles and have an
indeterminacy in their position; this indeterminacy is essentially the
nucleus--nucleus distance.  As a result, in order to be sure that the surface
of interest has all atoms \emph{above} it, the shape cannot be that of surface
A in Figure~\ref{mofig}.  It must be a flat plane that remains below the whole
mountain, like surface B in the figure.  However, a flat surface rising 
through a mountain does
not allow to exceed the mass change limit.

\csbul The multiple atom attempt.  One can imagine a number of atoms equal to
the number of the atoms of a mountain, but arranged in a way that all lie with
large spacing (roughly) in a single plane.  Again, the plane could move with
high speed.  However, also in this case the uncertainty in the atomic
positions makes it impossible to say that the mass flow limit has been
exceeded.

\csbul The multiple black hole attempt.  Black holes are typically large and
their uncertainty in position is thus negligible.  The mass limit $c^3/4G$ or
power limit $c^5/4G$ correspond to the flow of a single black hole moving
through a plane at the speed of light.  Several black holes crossing a plane
together at just under the speed of light thus seem to beat the limit.
However, the surface has to be physical: an observer must be possible one each
of its points.  But no observer can cross a black hole.  A black hole thus
effectively punctures the moving plane surface; no black hole can ever be said
to cross a plane surface, even less so a multiplicity of black holes.  The
limit remains valid.

\csbul The multiple neutron star attempt.  The mass limit seems in reach when
several neutron stars (which are slightly less dense than a black hole of the
same mass) cross a plane surface at the same time, at high speed.  However,
when the speed approaches the speed of light, the crossing time for points far
from the neutron stars and for those that actually cross the stars differ by
large amounts.  Neutron stars that are almost black holes cannot be crossed in
a short time in units of a coordinate clock that is located far from the
stars. Again, the limit is not exceeded.

\csbul The boost attempt.  A boost can apparently be chosen in such a way that
a force value $F$ in one frame is transformed into any desired value
$F^\prime$ in the other frame.  However, this result is not physical.  To be
more concrete, one imagines a massive observer, measuring the value $F$, at
rest with respect to a large mass, and a second, primed observer moving
towards the large mass with relativistic speed, measuring the value
$F^\prime$.  Both observers can be thought to be as small as desired.  If one
transforms the force field at rest $F$ applying the Lorentz transformations,
the force $F^\prime$ for the moving observer can apparently reach extremely
high values, as long as the speed is high enough.  However, a force value must
be measured by an observer at the specific point.  One has thus to check what
happens when the rapid observer moves towards the region where the force would
exceed the force limit.  The primed observer has a mass $m$ and a radius $r$.
To be an observer, he must be larger than a black hole; in other words, its
radius must obey $r > 2Gm/c^2$, implying that the observer has a non-vanishing
size.  When the observer dives into the force field surrounding the sphere,
there will be an energy flow $E$ towards the observer given by the transformed
field value and the proper crossing area of the observer.  This interaction
energy can be made as small as desired, by choosing an observer as small as
desired, but it is never zero.  When the moving observer approaches the large
massive charge, the interaction energy increases continuously.  Whatever
choice for the smallness of the observer is made is not important.  Before the
primed observer arrives at the point were the force $F^\prime$ was supposed to
be much higher than the force limit, the interaction energy will reach the
horizon limits (\ref{bb191203}) or (\ref{b64745747}).  Therefore, a horizon
appears and the moving observer is prevented from observing anything at all,
in particular any value above the horizon force.

The same limitation appears when a charged observed tries to measure
electromagnetic forces, or when nuclear forces are measured.  In summary,
boosts do not help to beat the force limit.

\csbul The divergence argument.  In apparent contrast to what was said so far,
the force on a test mass $m$ at a radial distance $d$ from a Schwarzschild
black hole (for $\Lambda=0$) is 
given by (Ohanian \& Ruffini 1994)    
\begin{equation}
    F= \frac{GMm}{d^2\sqrt{1-\frac{2GM}{dc^2}}}\quad.
    \label{fgfgfgf}
\end{equation}
In addition, the inverse square law of universal gravitation states
that the force between to masses $m$ and $M$ is
\begin{equation}
    F= \frac{GMm}{d^2}\quad.
    \label{fgfgfgfa}
\end{equation}
Both expressions can take any value and suggest that no maximum force
limit exist.

A detailed investigation shows that the maximum force still holds.
Indeed, the force in the two situations diverges only for not physical
point-like masses.  However, the maximum force implies a minimum approach
distance to a mass $m$ given by
\begin{equation}
    d_{\rm min}= \frac{2Gm}{c^2}\quad.
\end{equation}
The minimum approach distance -- simplifying, this would be the corresponding
black hole radius -- makes it impossible to achieve zero distance between two
masses or between a horizon and a mass.  The finiteness of this length value
expresses that a mass can never be point-like, and that a (real) minimum
approach distance of $2Gm/c^2$ appears in nature, proportional to the mass.
If this minimum approach distance is introduced in equations (\ref{fgfgfgf})
and (\ref{fgfgfgfa}), one gets
\begin{equation}
    F= \frac{c^4}{4G}\, \frac{Mm}{(M+m)^2} \frac{1}{\sqrt{1-\frac{M}{M+m}}}
    \leqslant \frac{c^4}{4G} 
\end{equation}
and
\begin{equation}
    F= \frac{c^4}{4G}\, \frac{Mm}{(M+m)^2} \leqslant \frac{c^4}{4G} .
\end{equation}
The maximum force value is never exceeded.  Taking into account the size of
observers prevents exceeding the maximum force.

\csbul The wall attempt.  Force is momentum change.  For example, momentum
changes when a basketball is reflected from a large wall.  If many such balls
are reflected at the same time, it seems that a force on the wall larger than
the limit can be realized.  However, this is impossible.  Every wall has a tiny
surface gravity.  For a large, but finite number of balls, the energy flow
limit of the horizon equation (\ref{b64745747}) will be reached, thus
implying the appearance of a horizon.  In that case, no reflection is possible
any more, and again the force or power limit cannot be exceeded.

\csbul The classical radiation attempt.  It is also not possible to create a
force larger than the maximum force concentrating a large amount of light onto
a surface.  However, the same situation as for basketballs arises: when the
limit value $E/A$ given by the horizon equation (\ref{b64745747}) is reached,
a horizon appears that prevents breaking the limit.

\csbul The multiple lamp attempt.  It might seem possible to create a power
larger than the maximum power by combining two radiation sources that each
emit $3/4$ of the maximum value.  But also in this case, the horizon limit
(\ref{b64745747}) is achieved and thus a horizon appears that swallows the
light and prevents that the force or power limit is exceeded.  (The limited
lifetime of such lamps makes these horizons time-dependent.)

\csbul The electrical charge attempt.  One might try to get forces above the
limit by combining gravity and electromagnetism.  However, in this case, the
energy in the horizon equation, like the first law of black hole mechanics
(Wald 1993), only gets gets an additional term.  The energy is then a sum of
mass--energy and electromagnetic energy.  For example, in the simplest case,
that of a static and charged black hole, the energy $\delta E=c^2\delta m + V
\delta q$ crossing the horizon includes the product of the electrical
potential $V$ at the horizon and the amount of charge $q$ crossing the
horizon.  However, the maximum force and power values remain unchanged.  In
other words, electromagnetism cannot be used to exceed the force or power
limit.

\csbul The consistency argument.  If observers cannot be point-like, one might
question whether it is still correct to apply the original definition of
momentum change or energy change as the integral of values measured by
observers attached to a given surface.  In general relativity, observers
cannot be point-like, as seen above.  However, observers can be as small as
desired.  The original definition thus remains applicable when taken as a
limit procedure for an observer size that decreases towards zero.  Obviously,
if quantum theory is taken into account, this limit procedure comes to an end
at the Planck length.  This is not an issue for general relativity, as long as
the typical dimensions in the situation are much larger than this value.

\csbul The quantum attempt.  If quantum effects are neglected, it is possible
to construct surfaces with sharp angles or even fractal shapes that overcome
the force limit.  However, such surfaces are not physical, as they assume that
lengths smaller than the Planck length can be realized or measured.  The
condition that a surface be physical implies among others that it has an
intrinsic uncertainty given by the Planck length.  A detailed study shows that
quantum effects do not allow to exceed the horizon force.  The basic reason is
the mentioned equality of the maximum force with the quotient of the Planck
energy and the Planck length, both corrected by a factor of order one.  Since
both the Plan energy and the Planck length are limits in nature, quantum
effects do not help at overcoming the force or power limit.  (Schiller
1997-2004).

Similar results are found when any other Gedanken experiment is
imagined.  Discussing them is an interesting way to explore general
relativity. No Gedanken experiment is successful; in
all cases, horizons prevent that the maximum force is exceeded.
Observing a value larger than the force or power limit requires
observation across a horizon.  This is impossible.

Maximum force (or power) implies that point masses do not exist.  This
connection is essential to general relativity.  The habit of thinking with
point masses -- a remainder of Galilean physics -- is one of the two reasons
that the maximum force principle has remained hidden for more than 80 years.
The (incorrect) habit of believing that the proper size of a system can be
made as small as desired while keeping its mass constant avoids that the force
or power limit is noticed.  Many paradoxes around maximum force or power are
due to this incorrect habit.


To see the use of a maximum force or power for the exploration of gravity, one
can use a simple image.  Nature prevents large force values by the appearance
of horizons.  This statement can be translated in engineer's language.  To
produce a force or power requires an engine.  Every engine produces exhausts.
When the engine approaches the power limit, the mass of the exhausts is
necessarily so large that their gravity cannot be neglected.  The gravity of
the exhausts saturates the horizon equation and then prevents the engine from
reaching the force or power limit.

Force is change of momentum with time; power is change of energy with time.
Since both momentum and energy are conserved, all changes take place through a
boundary.  The force and power limit state that these values are upper limits
independently of the boundary that is used.  Even if the boundary surface is
taken to cross the whole universe, the observed momentum or energy change
through that surface is limited by the maximum values.  This requires a check
with experiments.

\section{Experimental data}

\noindent No experiment, whether microscopic -- such as particle collisions --
macroscopic, or astronomical, has ever measured force values near or even
larger than the limit.  Also the search for pace-time singularities, which
would allow to achieve the force limit, has not been successful.  In fact, all
force values ever measured are many magnitudes smaller than the maximum value.
This result is due to the lack of horizons in the environment of all
experiments performed so far.

Similarly, no power measurement has ever provided any exception to the power
limit.  Only the flow of energy through a horizon should saturate the power
limit.  Every star, gamma ray burster, supernova, galaxy, or galaxy cluster
observed up to now has a luminosity below $c^5/4G$.  Also the energy flow
through the night sky horizon is below the limit.  (More about this issue
below.)  The brightness of evaporating black holes in their final phase could
approach or equal the limit.  So far, none has ever been observed.  In the
same way, no counter-example to the mass change limit has ever been observed.
Finding any counter-example to the maximum force, luminosity or mass change
would have important consequences.  It would invalidate the present approach
and thus invalidate general relativity.

On the other hand, we have seen above that general relativity contains a
maximum force and power, so that every successful test of the field equations
underlines the validity of this approach.

The absence of horizons in everyday life is the {second} reason why the
maximum force principle has remained undiscovered for so long.  Experiments in
everyday life do not point out the force or power limit to explorers.  The
{first} reason why the principle remained hidden, as shown above, is the
incorrect habit of believing in massive point particles.  This is a
theoretical reason.  (Prejudices against the concept of force in general
relativity have also played a role.)  The principle of maximum force -- or of
maximum power -- has thus remained unnoticed for a long time because nature
hid it both from theorists and from experimentalists.

In short, past experiments do not contradict the limit values and do not
require or suggest an alternative theory of gravitation.  But neither does the
data directly confirm the limits, as horizons are rare in everyday life or in
accessible experimental situations.  The maximum speed at the basis of special
relativity is found almost everywhere; maximum force and maximum power are
found almost nowhere.  For example, the absence of horizons in particle
collisions is the reason that the force limit is not of (direct) importance in
this domain.

\section{Cosmological data}

A maximum power is the simplest possible explanation of Olbers' paradox.
Power and luminosity are two names for the same observable.  The sum of all
luminosities in the universe is finite; the light and all other energy emitted
by all stars, taken together, is finite.  If one assumes that the
universe is homogeneous and isotropic, the power limit $P \leq c^5/4G$ must be
valid across any plane that divides the universe into two halves.  The part of
the universes's luminosity that arrives on earth is then so small that the sky
is dark at night.  In fact, the actually measured luminosity is still smaller
than this estimate, since a large part of the power is not visible to the human
eye (since most of it is matter anyway).  In other words, the night is dark
because of nature's power limit.  This explanation is \emph{not} in contrast
to the usual one, which uses the finite lifetime of stars, their finite
density, their finite size, the finite age and the expansion of the universe.
In fact, the combination of all these usual arguments simply implies and
repeats in more complex words that the maximum power value cannot be exceeded.
However, this simple reduction of the traditional explanation seems unknown in
the literature.

A maximum force in nature, together with homogeneity and isotropy, implies
that the visible universe is of \emph{finite size}.  The opposite case would
be an infinitely large universe.  But in that case, any two halves of the
universe would attract each other with a force above the limit (provided the
age of the universe is sufficiently large).  The result can be made
quantitative by imagining a sphere whose centre lies at the earth, which
encompasses all the universe, and whose radius decreases with time almost as
rapidly as the speed of light.  The mass flow $dm/dt=\rho A v$ is predicted to
saturate the mass flow limit $c^3/4G$; thus one has
\begin{equation}
    \frac{dm}{dt} = \rho_{o} 4 \pi R^2_{o} c = \frac{c^3}{4G} \quad ,
\end{equation}
a relation also predicted by the Friedmann models.  The WMAP measurements
confirm that the present day total energy density $\rho_{o}$ (including dark
matter and dark energy) and the horizon radius $R_{o}$ just saturate the
limit value.  The maximum force limit thus predicts the observed size of the
universe.

In summary, so far, neither experiment nor theory has allowed to exceed the
maximum force and power values.  Nevertheless, the statement of a maximum
force given by $c^4/4G$ (and the corresponding maximum power) remains open to
experimental falsification.  Since the derivation of general relativity from
the maximum force or from the maximum power is now established, one can more
aptly call them \emph{horizon} force and \emph{horizon} power.

\section{Predictions}

\csbul A maximum force and power is equivalent to general relativity and thus
implies the inverse square law of gravitation for small speed and curvature
values.  A maximum force is \emph{not} equivalent to scalar-tensor theories or
to modifications of the universal law of gravitation.

\csbul The exploration of physical systems that are mathematical analogues of
black holes -- for example, silent (or acoustical) black holes, or optical
black holes -- should confirm the force and power limits.  Future experiments
in these domains might be able to confirm the horizon equations (\ref{b191203})
or (\ref{bb191203}) directly.

\csbul Another domain in which tests might be possible is the relation that
follows from maximum force for the measurement errors $\Delta E$ and $\Delta
x$.  (Schiller 1997-2004) 
For all physical systems one has
\begin{equation}
    \frac{\Delta E}{ \Delta x} \leq \frac{c^4}{4G} \quad.
\end{equation}
So far, all measurements comply with the relation.  In fact, the left side is
usually so much smaller than the right side that the relation is not
well-known.  To have a direct check, one must look for a system where a rough
equality is achieved.  This might be the case in binary pulsar systems.  Other
systems do not seem to allow checking the relation.  In particular, there does
not seem to be a possibility to test this limit in satellite laser ranging
experiments.  For example, for a position error of 1\,mm, the mass error is
predicted to be below $3\cdot 10^{23}$\,kg, which so far is always the case.

\csbul There is a power limit for all energy sources and energies.  In
particular, the luminosity of all gravitational sources is also limited by
$c^5/4G$.  Indeed, all formulas for gravitational wave emission contain this
value as upper limit (Ohanian \& Ruffini 1994). 
Similarly, all numerical relativity simulations, such as the power emitted
during the merger of two black holes, should never exceed the limit.

\csbul The night sky is a horizon.  The power limit, when applied to the night
sky, makes the testable prediction that the flow of all matter and radiation
through the night sky adds up exactly to the value $c^5/4G$.  If one adds the
flow of photons, baryons, neutrinos, electrons and the other leptons,
including any particles that might be still unknown, the power limit must be
precisely saturated.  If the limit is exceeded or not saturated, general
relativity is not correct.  Increasing the precision of this test is a
challenge for future investigations.

\csbul It might be that one day the amount of matter and energy falling into
some black hole, such as the one at the centre of the Milky Way, might be
measured.  If that is the case, the mass rate limit $dm/dt \leq c^3/4G$ could
be tested directly.

\csbul Perfectly plane waves do not exist in nature.  Neither electrodynamic
nor gravitational waves can be infinite in extension, as such waves would
carry more momentum per time through a plane surface than allowed by the force
limit.  Taken the other way round, a wave whose integrated intensity
approaches the force limit cannot be plane.  The power limit thus implies a
limit on the product of intensity $I$ (given as energy per time and area) and
curvature radius $R$ of the front of a wave moving with the speed of
light~$c$:
\begin{equation}
    4 \pi R^2 I \leq \frac{c^5}{4G} \quad.
    \label{intli}
\end{equation}
This statement is difficult to check experimentally, whatever the frequency
and type of wave might be, as the value appearing on the right hand side is
extremely large.  Possibly, future experiments with gravitational wave
detectors, X-ray detectors, gamma ray detectors, radio receivers or particle
detectors will allow testing relation (\ref{intli}) with precision.  In
particular, the non-existence of plane gravitational waves also excludes the
predicted production of singularities in case that two plane waves collide.

\csbul Since the maximum force and power limits apply to all horizons, it is
impossible to squeeze mass into smaller regions of space than those given by a
region completely limited by a horizon.  As a result, a body cannot be denser
than a (uncharged, non-rotating) black hole of the same mass.  Both the force
and power limits thus confirm the Penrose inequality.  The limits also provide
a strong point for the validity of cosmic censorship.

\csbul The power limit implies that the highest luminosity is only achieved
when systems emit energy at the speed of light.  Indeed, the maximum emitted
power is only achieved when all matter is radiated away as rapidly as
possible: the emitted power $P=Mc^2/(R/v)$ cannot reach the maximum value if
the body radius $R$ is larger than a black hole (the densest bodies of a given
mass) or the emission speed $v$ is lower than that of light.  The sources with
highest luminosity must therefore be of maximum density and emit entities
without rest mass, such as gravitational waves, electromagnetic waves or
(maybe) gluons.  Candidates to achieve the limit are bright astrophysical
sources as well as black holes in evaporation or undergoing mergers.

\section{Outlook}

In summary, the maximum force principle (or the equivalent maximum power
principle) was shown to allow a simple axiomatic formulation of general
relativity: the horizon force $c^4/4G$ and the horizon power $c^5/4G$ are the
highest possible force and power values.  General relativity follows from
these limits.  All known experimental data is consistent with the limits.
Moreover, the limits imply the darkness at night and the finiteness of the
universe.

It is hoped that the maximum force principle will have applications for the
teaching of the field.  The principle might bring general relativity to the
level of first year university students; only the concepts of maximum force,
horizon and curvature are necessary.

Apart from suggesting some experimental tests, the principle of maximum force
also provides a guide for the search of a unified theory of nature that
incorporates general relativity and quantum theory.  Any unified theory of
nature must state that the value $c^4/4G$ is a maximum force.  Within an
uncertain numerical factor, this is the case for string theory, where a
maximum force appears, the so-called {string tension}.  A maximum force is
also predicted by loop quantum gravity.  Both string theory and loop quantum
gravity thus do predict gravity, as long as the predicted maximum force has
the correct value.

\section*{References}

\def\bibitem#1{\item}
\begin{list}{}{\parindent 0em \leftmargin 0em}

\bibitem{gibb} 
Gibbons, G.W., The maximum tension principle in general
relativity, {\it Foundations of Physics} {\bf 32}, {1891-1901} ({2002}).

\bibitem{sprelbook} D'Inverno, R., {\it Introducing Einstein's Relativity},
(Clarendon Press, 1992) page 36.

\bibitem{jac} Jacobson, T., Thermodynamics of spacetime: the Einstein equation
of state, {\it Physical Review Letters} {\bf 75}, 1260-1263 ({1995}).

\bibitem{textbook} {Ohanian, H.C. \& Ruffini, R.}, {\it Gravitation and
Spacetime} (W.W. Norton \& Co., {1994}).

\bibitem{sprelbook} Rindler, W., {\it Relativity, Special,
General and Cosmological} (Oxford University Press, 2001), page 70.

\bibitem{me20} {Schiller}, C., {\it Motion Mountain -- A Hike Beyond
Space and Time Along the Concepts of Modern Physics}
({{http://www.motionmountain.net}}, {1997-2004}), {section 7:
Maximum force -- a simple principle encompassing general relativity}.

\bibitem{me19} {Schiller}, C., {\it Motion Mountain -- A Hike Beyond
Space and Time Along the Concepts of Modern Physics}
({{http://www.motionmountain.net}}, {1997-2004}), {section 36:
Maximum force and minimum distance -- physics in limit statements}.

\bibitem{wald} {Wald}, R.M., {\it The first law of black hole mechanics}, in
B.L. Hu, M.P. Ryan, C.V. Vishveshwara, {\it Directions in General Relativity},
Proceedings of the 1993 International Symposium, Maryland: Papers in Honor of
Charles Misner, (Cambridge University Press, 1993).

\end{list}

\end{document}